\documentclass{moriond}
\usepackage{amsmath}

\def\be{\begin{equation}}
\def\ee{\end{equation}}
\def\bea{\begin{eqnarray}}
\def\eea{\end{eqnarray}}

\newcommand{\Photo}{}

\begin{document}
\vspace*{4cm}
\title{$n-n'$ OSCILLATIONS: SENSITIVITY OF A FIRST UCN BEAM EXPERIMENT}

\author{ W. SAENZ-AREVALO$^{\textrm{a}}$ on behalf of the collaboration}
\author{G. Ban$^{\textrm{a}}$, J. Chen$^{\textrm{a}}$, P.-J. Chiu$^{\textrm{b}}$, B. Clément$^{\textrm{c}}$, M. Guigue$^{\textrm{d}}$, T. Jenke$^{\textrm{e}}$, P. Larue$^{\textrm{c}}$, T. Lefort$^{\textrm{a}}$, O. Naviliat-Cuncic$^{\textrm{a,f}}$, B. Perriolat$^{\textrm{e}}$, G. Pignol$^{\textrm{c}}$, S. Roccia$^{\textrm{c,e}}$, P. Schmidt-Wellenburg$^{\textrm{g}}$}

\address{$^{\textrm{a}}$Laboratoire de Physique Corpusculaire, Caen, France; $^{\textrm{b}}$University of Zurich, Zurich, Switzerland; $^{\textrm{c}}$Laboratoire de Physique Subatomique et de Cosmologie, Grenoble, France; $^{\textrm{d}}$Laboratoire de Physique Nucléaire et des Hautes Énergies, Paris, France; $^{\textrm{e}}$Institut Laue-Langevin, Grenoble, France;  $^{\textrm{f}}$Michigan State University, Michigan, United States; $^{\textrm{g}}$Paul Scherrer Institute, Villigen, Switzerland}

\maketitle\abstracts{Oscillations of the neutron into a hidden sector particle are processes predicted in various Standard Model extensions. This extra channel for neutron disappearance has not been tested experimentally in large portions of the oscillation parameter space. Several efforts have been recently made on revising the oscillation time limits at low mass-splitting in ultra-cold neutron (UCN) storage experiments, and at larger mass-splitting in passing-through-wall experiments. In this work, we present the expected sensitivity of an experiment searching for neutron hidden neutron oscillations at intermediate mass-splitting via the application of magnetic fields in the range $B_0=30-1100$ $\mu$T. This experiment was performed at the Institut-Laue-Langevin using a novel UCN counter to monitor the beam flux. The measured UCN rate and the data collection technique predict a sensitivity on the oscillation time at the level of a couple of seconds.}


\section{Introduction}

Models predicting the existence of hidden matter have gained special interest since they address several yet unsolved problems in particle physics and cosmology simultaneously \cite{foot}. One of these models presents hidden matter as an exact copy of all known particles and their interactions, named mirror matter. Particles in this extra sector, which were originally proposed as a solution to the P-violation observed in the weak interaction \cite{lee}, could mix with ordinary particles through non-standard model interactions \cite{sarrazin}. Such processes could lead to new channels of CP and baryon symmetries violation, and therefore shed light on the baryogenesis problem \cite{ber16}. In addition, it has been argued that both sectors also interact through gravity, allowing to present mirror matter as a natural candidate for dark matter.

The mixing of the neutron ($n$) with its hidden counterpart ($n'$) is described \footnote{Using natural units $\hbar=c=1$} by the Hamiltonian \cite{hoster}
\begin{equation}
\hat{H}=\begin{pmatrix}
m_n + \Delta E & \epsilon_{nn'} \\
\epsilon_{nn'} & m_n + \delta m
\end{pmatrix},
\end{equation}
with $\Delta E$ the neutron energy due to its interaction with environment, $\delta m= m_{n'}-m_n$ the mass-splitting between the neutron and the hidden neutron masses, and $\epsilon_{nn'}=1/\tau_{nn'}$ the mixing parameter accounting for the briefness of the oscillations. The general purpose of $n-n'$ probing experiments is to trigger the $n-n'$ oscillations by canceling the energy degeneracy, i.e. by matching $\Delta E$ and $\delta m$. This has been done via small magnetic fields ($\Delta E=\mu_n B_0\sim10^{-3}$ neV) in UCN storage experiments \cite{abel}, and through material Fermi potentials ($\Delta E = V_F\sim 10^2$ neV) in regeneration experiments \cite{stereo}. Apart from the 5$\sigma$ anomaly reported by Berezhiani \cite{ber12} in UCN storage experiments, no confirmation of $n-n'$ oscillations has been reported so far. The present work targets scanning $n-n'$ oscillations on the yet unexplored energy range $\Delta E = 10^{-3}-10^{-1}$ neV via magnetic fields in the range $B_0 = 30-1100$ $\mu$T. Moreover, inspection of this energy range permits the reexamination of the intervals in the mass-splitting favored by the 5$\sigma$ anomaly. 
\section{Experimental setup}
\begin{figure}[t!]
\centerline{\includegraphics[width=.85\linewidth]{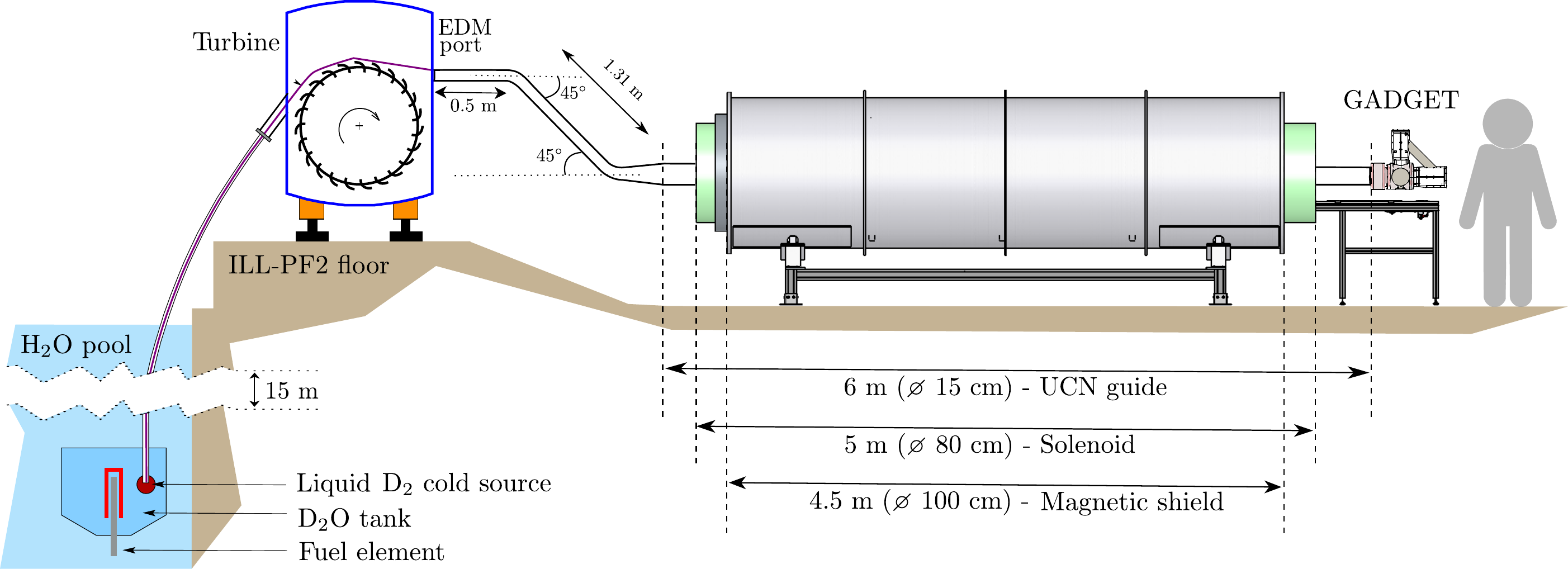}}
\caption{UCN beam setup for testing $n-n'$ oscillations at the PF2/EDM beam port at ILL reactor. }
\label{fig:setup}
\end{figure}
At the EDM port of the PF2 instrument at the ILL we studied $n-n'$ oscillations by direct measurement of the UCN flux. Unlike past UCN experiments, the setup did not include a storage chamber. Instead, we searched for oscillations during the UCN passage across a 6-m-long guide connecting the EDM beam port to an UCN counter. In order to scan the $\delta m$ parameter in the targeted range, a static magnetic field generated by a 5-m-long solenoid is applied along the main guide (Figure \ref{fig:setup}). If the degeneracy cancels for a given magnetic field ($|\delta m|=\mu_nB_0$), a signal of $n-n'$ oscillations would be observed as a drop of the detected beam flux. Two correction coils were added at the solenoid edges to extend the magnetic-field coverage and increase its uniformity. In addition, a 4.5-m-long cylindrical mu-metal shield was placed around the solenoid to improve the magnetic-field uniformity.

\section{UCN counting and background analysis}

For the measurement of the UCN beam flux we used a novel gaseous UCN detector (GADGET). This fast counter, specially developed by us for the n2EDM project \cite{n2edm}, was essential at the ILL EDM beam port where UCN fluxes can reach up to $10^6$ s$^{-1}$. The fast performance of GADGET is based on its detection principle. After crossing a 30-$\mu$m-thick entrance window, UCNs enter the detector sensitive volume filled with a gas-mixture of $^3$He and CF$_4$ at 15 and 500 mbar, respectively. Following the UCN absorption by the $^3$He nuclei, the emitted reaction products, $^1$H and $^3$H, induce scintillation on the CF$_4$ molecules whose decay times are of 6 ns \cite{lehaut}. A set of three photo-multiplier tubes (PMT) optically coupled to the gas chamber convert the scintillated light into voltage pulses lasting up to 30 ns. These signals were read and filtered with a triple coincidence window by the FASTER digital acquisition system, capable of sampling at 500 MHz with 12 bits. 

\begin{figure}[t!]
\centerline{\includegraphics[width=.9\linewidth]{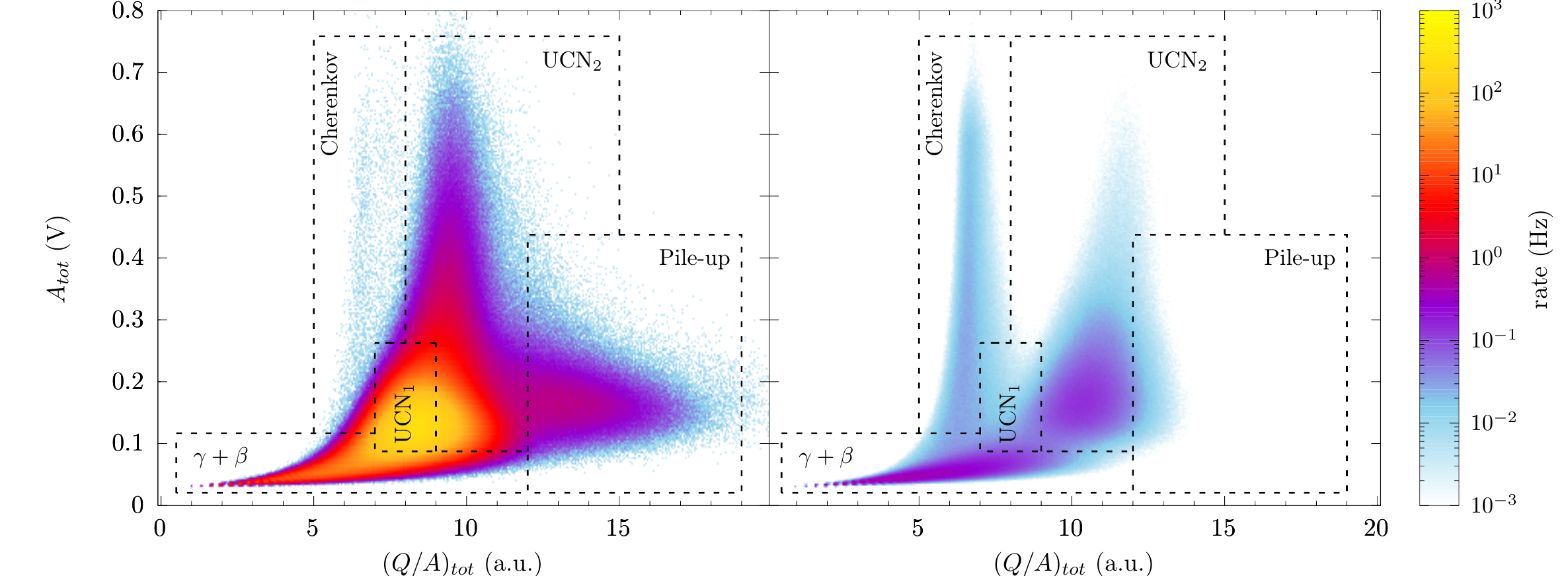}}
\caption{Pulse shape analysis from GADGET counter measuring in normal conditions (left) and with no $^3$He gas absorber (right). The total counting rates are 280 kHz and 1 kHz, respectively.}
\label{fig:psa}
\end{figure}

Background discrimination was based on pulse shape analysis (PSA) constructed with the total integrated charge ($Q_{tot}$) and amplitude ($A_{tot}$) over the three PMT signals. Figure \ref{fig:psa} shows the PSA maps measured for two detector configurations during UCN delivery: 200 s of ordinary detection (left side) and 3 hours recording with only CF$_4$ gas filling at 500 mbar, i.e. no $^3$He (right side). This last configuration allows the identification of three light-emitting processes different from the main UCN detection. First, events produced from $\gamma$ and $\beta$ environmental backgrounds interacting with the CF$_4$ molecules are mostly contained in category `$\gamma+\beta$'. Second, events created from the $\gamma$ interaction with electrons of the chamber's quartz windows are enclosed by the `Cherenkov' category. Finally, ambiguous events also including UCN capture on nuclei different from $^3$He build up a region mainly contained by category `UCN$_2$'. This partitioning allows defining the `UCN$_1$' category essentially free of background. Taking into account the counting rates of both measured configurations, background contamination in `UCN$_1$' during ordinary detection is estimated to be 0.04$\%$, which is 5 times smaller than the statistical fluctuations.


\section{Data taking and expected sensitivity}

We formulate a self-normalized measurement of the UCN by counting the number of neutrons ($N$) at three magnetic fields $A$, $B$ and $C$. Performing a double counting at $B$, the normalized ratio sensitive to $n-n'$ oscillations was defined as
\begin{equation}
    R_{ABC}=\frac{N_B+N'_B}{N_A+N_C} \left\{\begin{array}{lc}
        = 1, & \textrm{if no oscillations}    \\
        < 1, & \textrm{if oscillations at field } B \\
        > 1, & \textrm{if oscillations at field } A \textrm{ or } C.
        \end{array}\right.
        \label{eq:R}
\end{equation}
Single measurements of $R_{ABC}$ were obtained from the counting of UCN passing through the experimental setup in cycles lasting 200 s. Few seconds of every cycle were discarded to ramp the magnetic field and the rest was divided into 4 batches of 44 s where the magnetic field values followed the sequence
\begin{equation}
    \{A,B,B,C\}=\{B_0-20\,\mu\textrm{T},B_0,B_0,B_0+20\,\mu\textrm{T}\}.
\end{equation}
The 20 $\mu$T magnetic field step inside a cycle was chosen larger than the resonance width ($\sim 2.4$ $\mu$T) so that oscillations would occur at only one of the field configurations. Correspondingly, the magnetic field shifting step ($\Delta B_0$) in between cycles was set in $\Delta B_0=3$ $\mu$T to guarantee that all the intermediate values were covered by the scanning process.

\begin{figure}[t!]
\centerline{\includegraphics[width=.8\linewidth]{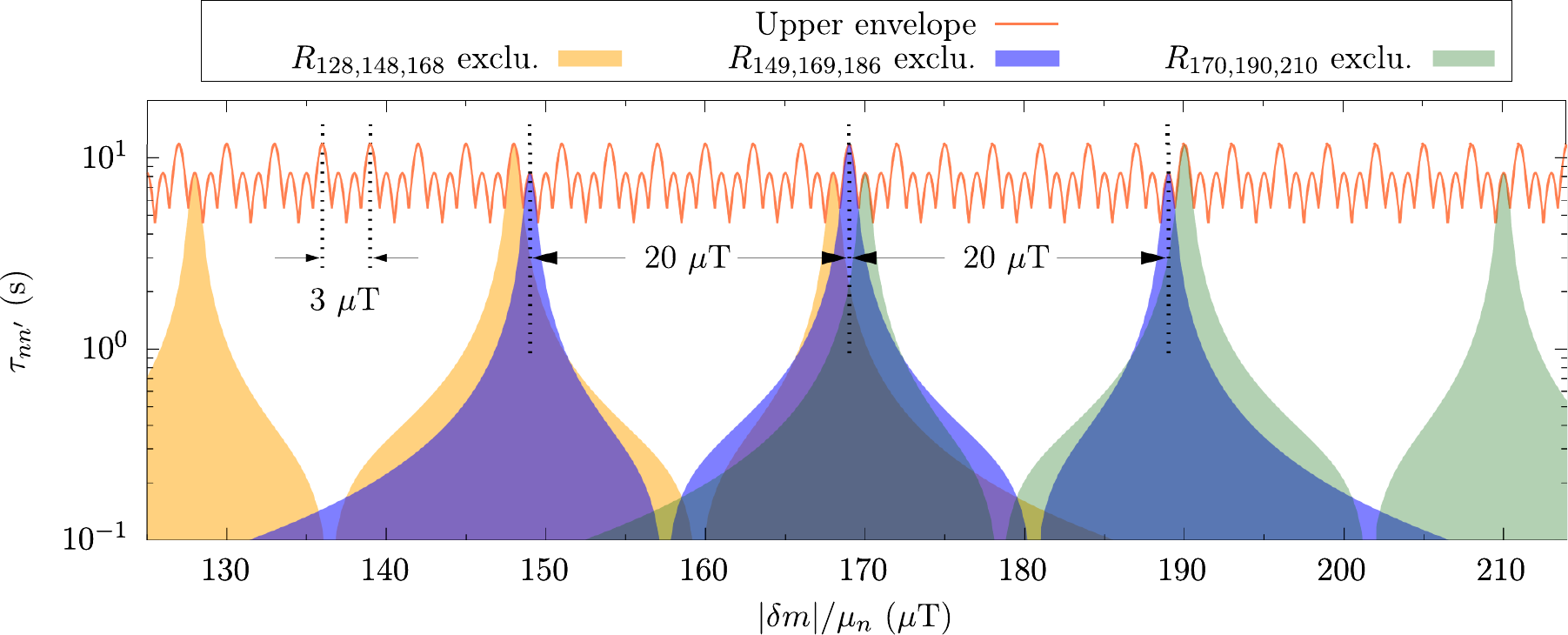}}
\caption{Expected sensitivity of the UCN beam experiment. The upper envelope is built with the 95\% C.L. exclusion regions between 125 and 215 $\mu$T. Individual contributions of three cycles are displayed with colored regions to better illustrate the upper envelope shape. The same pattern extends over the full $|\delta m|/\mu_n$ range.}
\label{fig:sensitivity}
\end{figure}

According to Eq. \ref{eq:R}, a $n-n'$ oscillation signal is identified if $|R_{\textrm{\tiny{$ABC$}}}-1|> \zeta \Delta R_{\textrm{\tiny{$ABC$}}}$, with $\zeta$ a coefficient depending on the confidence level (CL), and $\Delta R_{\textrm{\tiny{$ABC$}}}$ the uncertainty of $R_{\textrm{\tiny{$ABC$}}}$. In contrast, if all measurements of $|R_{\textrm{\tiny{$ABC$}}}-1|$ are contained within their uncertainties, the model parameters $\tau_{nn'}$ and $\delta m$ can be bounded. Figure \ref{fig:sensitivity} shows the exclusion regions expected from the `no-signal' scenario assuming $R_{\textrm{\tiny{$ABC$}}}=1$ and an UCN flux $\Phi_{\textrm{\tiny{UCN}}}\sim280$ kHz, with uncertainties purely derived from statistical fluctuations, i.e. $\Delta R_{\textrm{\tiny{$ABC$}}}\sim \sqrt{4/(\Phi_{\textrm{\tiny{UCN}}}\cdot44\,\textrm{s})}\sim 2.7\times10^{-4}$. Every cycle measurement determines an exclusion region characterized by three maxima located at $|\delta m|/\mu_n = A,B$ and $C$. The sensitivity at $B$ is two times larger than the one at $A$ or $C$ due to the larger counting time at the former ($2\times 44$ s). After scanning the triplet $ABC$ every 3 $\mu$T and overlapping the resulting exclusion regions, the upper envelope shows that $\tau_{nn'}>4$ s for any $|\delta m|/\mu_n$ contained in the targeted interval $[30-1100]$ $\mu$T.

\section{Outlook}
Given that no measurement has constrained $\delta m$ in the targeted interval at the scale of few seconds, this sensitivity analysis shows that the chosen scanning technique can result on a new experimental limit of $\tau_{nn'}$. A later publication will be dedicated to the actual exclusion obtained from the measured  $R_{\textrm{\tiny{$ABC$}}}$ points while considering extra model refinements such as the field uniformities, UCN velocity distributions, and fluctuations of the neutron flux.
\section*{Acknowledgments}
We would like to acknowledge the technical support by T. Brenner during the experiment.
\section*{References}

\end{document}